\documentclass[10pt,journal,final,twocolumn]{IEEEtran}

\usepackage{graphicx}
\usepackage{subfigure}
\usepackage{epstopdf}
\usepackage{epsfig}
\usepackage[cmex10]{amsmath}
\usepackage{amssymb}
\usepackage{times}
\usepackage{ntheorem}
\usepackage{pifont}
\usepackage{stfloats}
\usepackage{bm}
\usepackage{color}
\usepackage{makecell}
\usepackage{array}
\usepackage{multirow}

\begin{document}
%
\title{\huge Deep Learning based Channel Estimation for Massive MIMO with Mixed-Resolution ADCs}

\author{\IEEEauthorblockN{
Shen~Gao, Peihao~Dong,~\IEEEmembership{Student Member,~IEEE}, Zhiwen~Pan,~\IEEEmembership{Member,~IEEE},\\
and Geoffrey Ye Li,~\IEEEmembership{Fellow,~IEEE}
}

\vspace{-0.5cm}
\thanks{
The work of S. Gao and Z. Pan was supported in part by the National Key Research and Development Project under Grant 2018YFB1802402. The work of S. Gao was also supported by the China Scholarship Council (CSC) under Grant 201706090065.

S. Gao and Z. Pan are with the National Mobile Communications Research Laboratory, Southeast University, Nanjing 210096, China, and are also with the Purple Mountain Laboratories, Nanjing, China (e-mail: gaoshen@seu.edu.cn; pzw@seu.edu.cn).

P. Dong is with the National Mobile Communications Research Laboratory, Southeast University, Nanjing 210096, China (e-mail: phdong@seu.edu.cn).

G. Y. Li is with the School of Electrical and Computer Engineering, Georgia Institute of Technology, Atlanta, GA 30332 USA (e-mail: liye@ece.gatech.edu).
}
\vspace{-0.4cm}
}

\IEEEtitleabstractindextext{%
\begin{abstract}
In this article, deep learning is applied to estimate the uplink channels for mixed analog-to-digital converters (ADCs) massive multiple-input multiple-output (MIMO) systems, where a portion of antennas are equipped with high-resolution ADCs while others employ low-resolution ones at the base station. A direct-input deep neural network (DI-DNN) is first proposed to estimate channels by using the received signals of all antennas. To eliminate the adverse impact of the coarsely quantized signals, a selective-input prediction DNN (SIP-DNN) is developed, where only the signals received by the high-resolution ADC antennas are exploited to predict the channels of other antennas as well as to estimate their own channels. Numerical results show the superiority of the proposed DNN based approaches over the existing methods, especially with mixed one-bit ADCs, and the effectiveness of the proposed approaches on different ADC resolution patterns.
\end{abstract}

}

\maketitle

\IEEEdisplaynontitleabstractindextext

%
\IEEEpeerreviewmaketitle

\vspace{-0.2cm}
\section{Introduction}

\IEEEPARstart{M}{assive} multiple-input multiple-output (MIMO) is a promising technique to achieve high transmission rate and continuous wide coverage for future cellular communication systems \cite{L. Lu}$-$\cite{L. Dai}. In massive MIMO, the hardware cost and power consumption are the main obstacles to equipping the large-scale array with the dedicated radio frequency (RF) chain for each antenna at the base station (BS). To address this problem, low-resolution analog-to-digital converters (ADCs), e.g. one to three bits, are employed to replace the power-hungry and expensive high-resolution ADCs \cite{L. Fan}, \cite{P. Dong_a}. However, the low-resolution ADCs incur the severely nonlinear distortion, which poses daunting challenges in channel estimation and data detection.

Mixed-ADC is proposed in \cite{N. Liang} to achieve a good tradeoff between the cost and the performance. It has also been revealed by \cite{J. Zhang_a} that the mixed-ADC architecture can reduce the channel estimation overhead and signal processing complexity remarkably. In addition, mixed-ADC also guarantees the backward compatibility with current cellular networks by directly placing antennas with low-resolution ADCs at the BS. The closed-form approximation of the achievable rate is derived in \cite{J. Zhang} for the mixed-ADC massive MIMO uplink under Rician fading channels. For the massive MIMO uplink with different ADC resolution levels at different antennas, the approximation of the outage probability is developed in \cite{Q. Ding}. The sum rate performance of the multi-user massive MIMO relaying system with mixed-ADC at the BS is analyzed in \cite{J. Liu}. \cite{J. Zhang_c} considers mixed-ADC/digital-to-analog converters (DACs) at the massive MIMO relay and derives achievable rate expressions for further system optimization.

Although the mixed-ADC can reduce hardware cost and signal processing complexity significantly, it still limits the performance of the tranceiver, especially for mixed one-bit ADC. Machine learning (ML) has been recently introduced into wireless communications and achieved attractive success by extracting the inherent correlation from data \cite{Z.-J. Qin}, \cite{H. He_a}. It is a potential way to efficiently improve the performance with mixed-ADC. ML algorithms have been used to address the intractable problems that the traditional methods are unable to handle well. In \cite{H. Ye}, deep learning (DL) is successfully used in joint channel estimation and signal detection with interference and non-linear distortions. In \cite{P. Dong_b}, channel correlation is exploited by deep convolutional neural network (CNN) to improve the channel estimation accuracy and to reduce the computational complexity for millimeter wave massive MIMO systems. In \cite{Y.-S. Jeon}, a supervised learning based successive interference cancellation is developed for MIMO detection with low-resolution ADCs. More results on DL for physical layer communications can be found in \cite{Z.-J. Qin}.

In this article, we propose the DL based channel estimation framework for mixed-ADC massive MIMO uplink, which to our knowledge has not been considered in literature. The novelty and contribution of this article can be summarized as follows:
\begin{itemize}[\IEEEsetlabelwidth{Z}]
\item[1)] We first propose the direct-input deep neural network (DI-DNN) based approach by simply inputting the signals of all antennas into a fully-connected DNN. To eliminate the adverse impact of the severely distorted signals quantized by the low-resolution ADCs on estimation accuracy, we propose a novel prediction mapping from the channels of high-resolution ADC antennas to those of low-resolution ADC antennas. The inherent nonlinear relationship between them is intractable to be modeled by the traditional methods but can be well captured by the DNN and thus we develop the selective-input prediction (SIP-DNN) based approach to implement it. This novelty coincides with the prior works that exploit the mature DNN to address the intractable problems that the traditional methods are unable to handle well \cite{H. Ye}, \cite{P. Dong_b}.

\item[2)] Numerical results show that the proposed DNN based channel estimation approaches outperform the state of the art methods and can be applied to different ADC resolution patterns in practical systems. We also find that the DI-DNN is suitable to the case with fewer high-resolution ADC antennas or low signal-to-noise ratio (SNR) while the SIP-DNN is preferable otherwise.
\end{itemize}

\emph{Notations}: In this article, we use upper and lower case boldface letters to denote matrices and vectors, respectively. $\lVert\cdot\rVert$, $(\cdot)^T$, and $\mathbb{E}\{\cdot\}$ represent the Euclidean norm, transpose, and expectation, respectively. $|\mathcal{X}|$ denotes the cardinality of set $\mathcal{X}$. $\mathcal{CN}(\mu,\sigma^2)$ represents circular symmetric complex Gaussian distribution with mean $\mu$ and variance $\sigma^2$. $\textrm{Re}(\cdot)$ and $\textrm{Im}(\cdot)$ denote the real part and imaginary part, respectively.

\vspace{-0.2cm}
\section{System Model}

\begin{figure}[t]
\centering
\includegraphics[trim=15 10 0 0, width=2.8in]{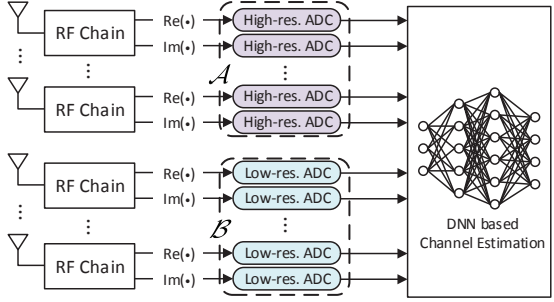}
\caption{System model of a massive MIMO uplink with mixed-ADC.}\label{System_model}
\vspace{-0.4cm}
\end{figure}

As shown in Fig.~\ref{System_model}, we consider a mixed-ADC massive MIMO uplink, where $M$ antennas at the BS are divided into two sets with high- and low-resolution ADCs, respectively. Denote $\mathcal{A}$ and $\mathcal{B}$ as the index sets of the antennas with high- and low-resolution ADCs, respectively, where $\mathcal{A}\cap\mathcal{B}=\emptyset$ and $\mathcal{A}\cup\mathcal{B}=\{1,2,\ldots,M\}$. Then we have $|\mathcal{A}|+|\mathcal{B}|=M$ and denote $\eta=\frac{|\mathcal{A}|}{M}$ as the ratio of the number of high-resolution ADC antennas over the total number of antennas. One single-antenna user is considered for simplicity even if the proposed approaches can be also applied to multi-user case. The multi-path channels from the user to the BS is given by \cite{H. Q. Ngo_b}
\vspace{-0.2cm}
\begin{equation}
\label{eqn_h}
\mathbf{h}=\sum_{l=1}^L \alpha_{l}\mathbf{a}(\varphi_l),
\end{equation}
where $L$ is the number of paths and $\alpha_{l}\sim \mathcal{CN}(0, \sigma_{\alpha}^2)$ is the propagation gain of the $l$th path with $\sigma_{\alpha}^2$ being the average power gain. For the $l$th path, $\varphi_l$ is the azimuth angles of arrival (AoA) at the BS and $\mathbf{a}(\varphi_l)$ denotes the corresponding response vector. For a uniform linear array, $\mathbf{a}(\varphi_l)$ is given by
\vspace{-0.2cm}
\begin{equation}
\label{eqn_au}
\mathbf{a}(\varphi_l)\!=\!\frac{1}{\sqrt{L}}\bigl[1,e^{-j2\pi\frac{d}{\lambda}\!\sin(\varphi_l)},\ldots,e^{-j2\pi\frac{d}{\lambda}(M\!-\!1)\!\sin(\varphi_l)}\bigr]^{T}\!,\!
\end{equation}
where $d$ denotes the space between the adjacent antennas at the BS and $\lambda$ is the wavelength of the carrier frequency. The antennas in $\mathcal{A}$ and $\mathcal{B}$ are placed at the BS with some pattern with $\mathbf{h}_{\mathcal{A}}$ and $\mathbf{h}_{\mathcal{B}}$ denoting the channels from the user to the antennas in $\mathcal{A}$ and $\mathcal{B}$, respectively.

\vspace{-0.2cm}
\section{DNN based Channel Estimation}

\begin{figure}[!t]
  \centering
  \subfigure[\scriptsize DI-DNN based approach]{\includegraphics[trim=0 0 0 0, width=1.6in]{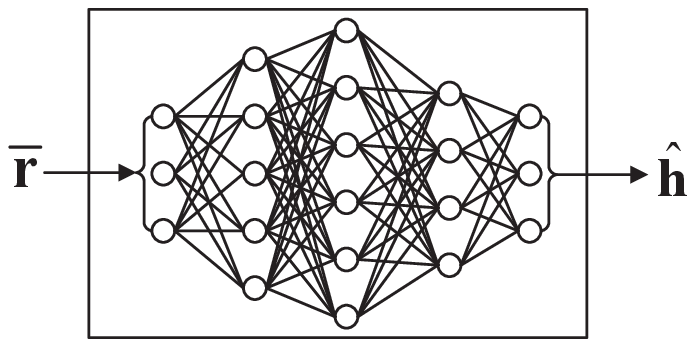}\label{DI_DNN}}
  \vspace{0.1cm}
  \subfigure[\scriptsize SIP-DNN based approach]{\includegraphics[width=3.1in]{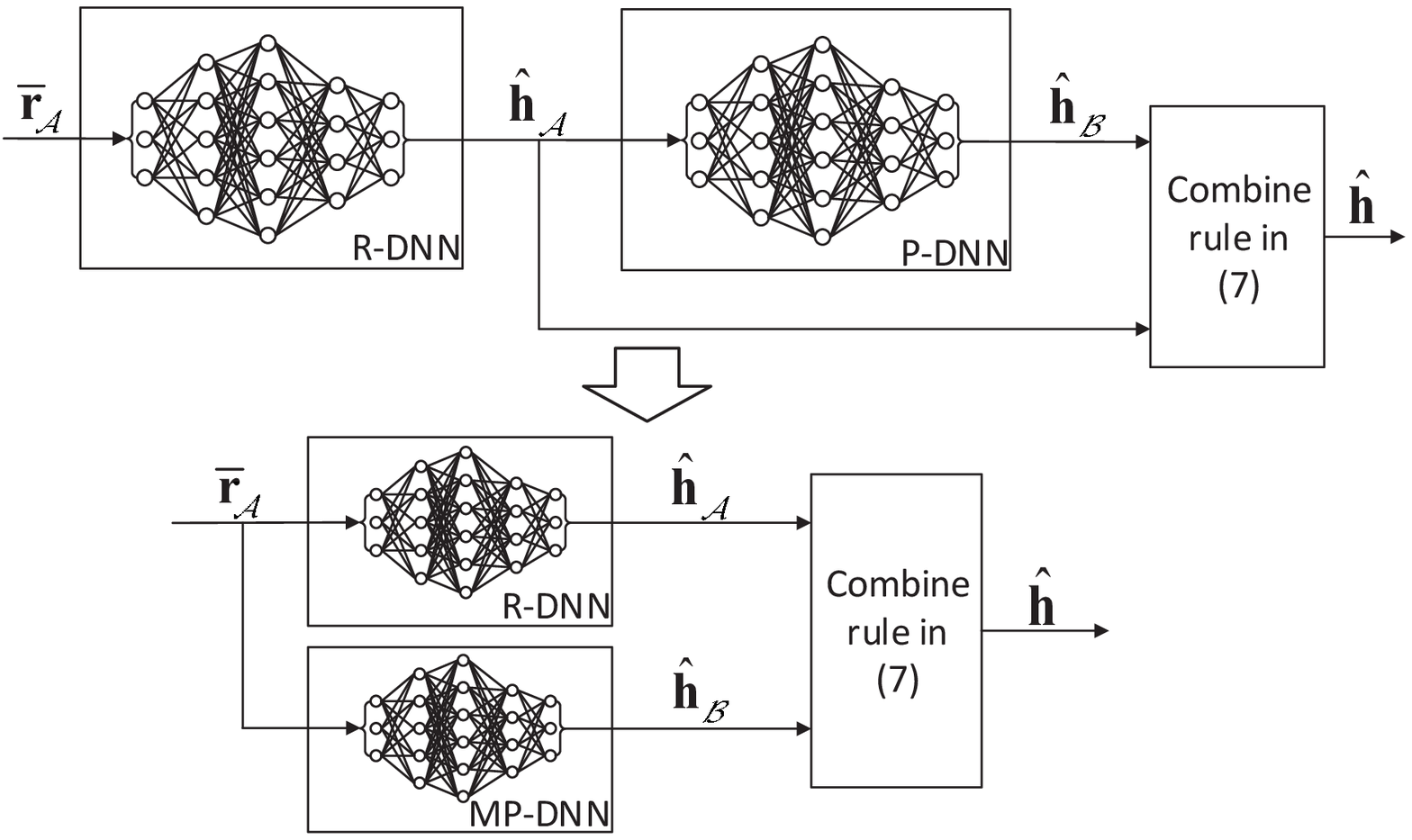}\label{SIP_DNN}}
  \caption{DNN based channel estimation.}\label{DNN}
  \vspace{-0.6cm}
\end{figure}

To estimate the uplink channels, the user transmits a pilot signal, $\sqrt{P}x$, to the BS with $P$ denoting the transmit power. Without loss of generality, we set $x=1$ since it is known to the BS. Then the received pilots at the BS is written as
\begin{equation}
\label{eqn_y}
\mathbf{y}=\sqrt{P}\mathbf{h}+\mathbf{n},
\end{equation}
where $\mathbf{n}$ denotes the additive white Gaussian noise vector at the BS with independent and identically distributed $\mathcal{CN}(0,\sigma_0^2)$ elements.

Then the elements of $\mathbf{y}$ corresponding to $\mathcal{A}$ and $\mathcal{B}$ are quantized by the high- and low-resolution uniform quantizers, respectively, and the $m$th element of signals input into the baseband processor for channel estimation is given by
\begin{equation}
\label{eqn_r}
\left[\mathbf{r}\right]_{m}=
\begin{cases}
\left[\mathbf{y}\right]_{m}, & m\in \mathcal{A},\\
\mathbb{Q}\left(\left[\mathbf{y}\right]_{m}\right), & m\in \mathcal{B},
\end{cases}
\end{equation}
where $\mathbb{Q}\left(\cdot\right)$ denotes the element-wise quantization operation on the real and imaginary parts separately. In (\ref{eqn_r}), we have ignored the quantization error of the high-resolution ADCs in antenna set $\mathcal{A}$. Then the least square (LS) estimation of $\mathbf{h}$ is given by $\bar{\mathbf{r}}=\frac{\mathbf{r}}{\sqrt{P}}$, which can be divided into two sub-vectors, $\bar{\mathbf{r}}_{\mathcal{A}}$ and $\bar{\mathbf{r}}_{\mathcal{B}}$, for $\mathcal{A}$ and $\mathcal{B}$, respectively.

\vspace{-0.2cm}
\subsection{DI-DNN based Approach}

The coarsely quantized output of $\mathcal{B}$ with low-resolution ADCs hinder the corresponding channel estimation. Fortunately, the limited number of scattering clusters in the propagation environment and finite physical space between antennas at the BS introduce the correlation among the received signal of each antenna. Then the almost undistorted quantization signals of $\mathcal{A}$ can provide additional channel information in spatial domain for $\mathcal{B}$ to improve the channel estimation accuracy.

Inspired by the aforementioned fact, we propose the DI-DNN based approach, where the LS estimation of $\mathcal{A}$ and $\mathcal{B}$ are simultaneously input into a DNN to predict the channels of all antennas as shown in Fig.~\ref{DI_DNN}. That is,
\begin{equation}
\label{eqn_FC_DNN}
\hat{\mathbf{h}}=f_{Q}\left(\mathbf{W}_{Q}f_{Q-1}\left(\cdots f_2\left(\mathbf{W}_{2}\bar{\mathbf{r}}\right)\right)\right),
\end{equation}
where $\hat{\mathbf{h}}$ is the estimation of $\mathbf{h}$, $Q$ is the total number of neural layers, $\mathbf{W}_{q}$ and $f_{q}(\cdot)$ denote the weight matrix and activation function of the $q$th layer $\forall q=2,\ldots,Q$. By setting the activation functions properly and updating weight matrices in the data-driven manner, the DI-DNN aims to minimize the MSE over all training samples, which is given by
\vspace{-0.1cm}
\begin{equation}
\begin{aligned}
\label{eqn_mse}
\textrm{MSE}=\frac{1}{N_{\textrm{tr}}}\sum_{n=1}^{N_{\textrm{tr}}}\left\|\mathbf{h}_{n}-\hat{\mathbf{h}}_{n}\right\|^2,
\end{aligned}
\end{equation}
where $N_{\textrm{tr}}$ denotes the number of training samples, $\mathbf{h}_{n}$ and $\hat{\mathbf{h}}_{n}$ denote the true channel and the channel approximated by the DI-DNN, respectively, for the $n$th training sample.

The DI-DNN consists of the input layer, three hidden layers with rectified linear unit (ReLU) activation function, and the output layer with hyperbolic tangent activation function. The $n$th training sample is denoted as $\left(\bar{\mathbf{r}}_{n},\frac{\mathbf{h}_{n}}{c}\right)$, where the input data $\bar{\mathbf{r}}_{n}$ is the LS estimation of the true channel, $\mathbf{h}_{n}$, and $\frac{\mathbf{h}_{n}}{c}$ denotes the target data that the DNN tries to approximate. $c>0$ is a scaling constant to make the range of all target data match the tangent activation function of the output layer. The approximated channel corresponding to the $n$th training sample is expressed as $\hat{\mathbf{h}}_{n}=c\bar{\mathbf{h}}_{n}$, where $\bar{\mathbf{h}}_{n}$ is the output of the DI-DNN. Then the MSE in (\ref{eqn_mse}) is minimized by using the backpropagation algorithm. In this article, we consider the case with $M=64$. The LS channel estimation $\bar{\mathbf{r}}_n\in\mathbb{C}^{64\times1}$ is first converted to a $128\times 1$ real-valued vector by vertically stacking its real and imaginary parts. The real-valued vector is successively processed by three dense layers with $160$, $200$, and $160$ neurons, respectively. Finally, the output layer outputs $[\textrm{Re}(\bar{\mathbf{h}}_{n}^T), \textrm{Im}(\bar{\mathbf{h}}_{n}^T)]^T\in\mathbb{R}^{128\times1}$ and then the approximated channel $\hat{\mathbf{h}}_{n}$ can be obtained. The detailed DNN architecture is summarized in Table~\ref{table_1}.

\begin{table}[!t]
\centering
\caption{Architecture of the DI-DNN}
\label{table_1}
\begin{tabular}{c|c|c}
\hline
 \makecell[tl]{Neural layer} & \makecell{Size} & Activation function\\
\hline
 Input layer & 128 & - \\
\hline
 Dense layer 1 & 160 & ReLU \\
\hline
 Dense layer 2 & 200 & ReLU \\
\hline
 Dense layer 3 & 160 & ReLU \\
\hline
 Output layer & 128 & tanh \\
\hline
\end{tabular}
\vspace{-0.3cm}
\end{table}

\vspace{-0.2cm}
\subsection{SIP-DNN based Approach}

The DI-DNN based approach proposed in Section III.A improves the channel estimation accuracy of $\mathcal{B}$ resorting to the undistorted signals of $\mathcal{A}$, but it ignores the adverse impact of the severely distorted quantization signals in $\mathcal{B}$ on the channel estimation performance, especially when $\mathcal{B}$ is with extremely low-resolution ADCs. To address this issue, the SIP-DNN based approach is developed in the following.

The basic idea of the SIP-DNN based approach can be summarized as follows: 1) Collect the reliable LS channel estimation of $\mathcal{A}$ and propose the prediction mapping from the channels in $\mathcal{A}$ to those in $\mathcal{B}$; 2) Utilize the DNN to implement this nonlinear mapping. The top figure of Fig.~\ref{SIP_DNN} illustrates the algorithm details of the SIP-DNN based approach. The LS channel estimation of $\mathcal{A}$, $\bar{\mathbf{r}}_{\mathcal{A}}$, is first refined by a refinement DNN (R-DNN), which outputs the more accurate estimation of $\mathbf{h}_{\mathcal{A}}$, i.e., $\hat{\mathbf{h}}_{\mathcal{A}}$. Then $\hat{\mathbf{h}}_{\mathcal{A}}$ is used to predict the channels corresponding to $\mathcal{B}$, i.e., $\hat{\mathbf{h}}_{\mathcal{B}}$, through a prediction DNN (P-DNN). Finally, $\hat{\mathbf{h}}_{\mathcal{A}}$ and $\hat{\mathbf{h}}_{\mathcal{B}}$ are combined
to obtain the full estimated channel, $\hat{\mathbf{h}}$, as
\begin{equation}
\label{eqn_h_AB}
\left[\hat{\mathbf{h}}\right]_{m}=
\begin{cases}
\left[\hat{\mathbf{h}}_{\mathcal{A}}\right]_{u_m}, & m\in \mathcal{A},\\
\left[\hat{\mathbf{h}}_{\mathcal{B}}\right]_{v_m}, & m\in \mathcal{B},
\end{cases}
\end{equation}
where $u_m$ and $v_m$ denote the indexes of the $m$th element of $\hat{\mathbf{h}}$ in $\hat{\mathbf{h}}_{\mathcal{A}}$ and $\hat{\mathbf{h}}_{\mathcal{B}}$ when $m\in \mathcal{A}$ and $\mathcal{B}$, respectively. The serial connection of R-DNN and P-DNN hinders the offline training and we propose to transform the SIP-DNN based approach to the bottom figure in Fig.~\ref{SIP_DNN}. In more details, the P-DNN is replaced by a modified prediction DNN (MP-DNN), which directly uses $\bar{\mathbf{r}}_{\mathcal{A}}$ instead of $\hat{\mathbf{h}}_{\mathcal{A}}$ to predict $\hat{\mathbf{h}}_{\mathcal{B}}$ and thus can be trained parallelly with the R-DNN. According to the simulation trails, this transformation will not cause performance loss but facilitate the offline training significantly.

For R-DNN and MP-DNN, we have
\begin{equation}
\label{eqn_FC_DNN1}
\hat{\mathbf{h}}_{\mathcal{A}}=g_{\textrm{R},S}\left(\mathbf{U}_{S}g_{\textrm{R},S-1}\left(\cdots g_{\textrm{R},2}\left(\mathbf{U}_{2}\bar{\mathbf{r}}_{\mathcal{A}}\right)\right)\right),
\end{equation}
\vspace{-0.2cm}
\begin{equation}
\label{eqn_FC_DNN2}
\hat{\mathbf{h}}_{\mathcal{B}}=g_{\textrm{MP},T}\left(\mathbf{V}_{T}g_{\textrm{MP},T-1}\left(\cdots g_{\textrm{MP},2}\left(\mathbf{V}_{2}\bar{\mathbf{r}}_{\mathcal{A}}\right)\right)\right),
\end{equation}
where $S$ and $T$ are the total numbers of neural layers, $g_{\textrm{R},s}(\cdot)$ and $g_{\textrm{MP},t}(\cdot)$ represent the activation function of the $s$th layer and $t$th layer, and $\mathbf{U}_{s}$ and $\mathbf{V}_{t}$ denote the corresponding weight matrices for R-DNN and MP-DNN, respectively, $\forall s=2,\ldots,S$ and $\forall t=2,\ldots,T$. The objectives of R-DNN and MP-DNN are to minimize their respective MSEs
\vspace{-0.2cm}
\begin{equation}
\begin{aligned}
\label{eqn_mse1}
\textrm{MSE}_{\textrm{R}}=\frac{1}{N_{\textrm{tr}}}\sum_{n=1}^{N_{\textrm{tr}}}\left\|\mathbf{h}_{\mathcal{A},n}-\hat{\mathbf{h}}_{\mathcal{A},n}\right\|^2,
\end{aligned}
\end{equation}
\vspace{-0.2cm}
\begin{equation}
\begin{aligned}
\label{eqn_mse2}
\textrm{MSE}_{\textrm{MP}}=\frac{1}{N_{\textrm{tr}}}\sum_{n=1}^{N_{\textrm{tr}}}\left\|\mathbf{h}_{\mathcal{B},n}-\hat{\mathbf{h}}_{\mathcal{B},n}\right\|^2,
\end{aligned}
\end{equation}
where $\mathbf{h}_{\Delta,n}$ and $\hat{\mathbf{h}}_{\Delta,n}$ denote the true channel of set $\Delta$ and the channel approximated by the corresponding DNN, respectively, with $\Delta=\mathcal{A}$ or $\mathcal{B}$ for the $n$th training sample.

Similar to the DI-DNN in Section III.A, both R-DNN and MP-DNN include the input layer, three hidden layers with ReLU activation function, and the output layer with hyperbolic tangent activation function. But they have different numbers of neurons and different weight matrices for each layer due to their respective tasks. For R-DNN, the $n$th training sample has the form of $\left(\bar{\mathbf{r}}_{\mathcal{A},n},\frac{\mathbf{h}_{\mathcal{A},n}}{c}\right)$, where the input data $\bar{\mathbf{r}}_{\mathcal{A},n}$ is the LS channel estimation of the true channel in $\mathcal{A}$, $\mathbf{h}_{\mathcal{A},n}$, and $\frac{\mathbf{h}_{\mathcal{A},n}}{c}$ denotes the target data that R-DNN tries to approximate. The approximated channel corresponding to $\mathcal{A}$ of the $n$th training sample is written as $\hat{\mathbf{h}}_{\mathcal{A},n}=c\bar{\mathbf{h}}_{\mathcal{A},n}$, where $\bar{\mathbf{h}}_{\mathcal{A},n}$ is the output of R-DNN. Then the MSE in (\ref{eqn_mse1}) is minimized by using the backpropagation algorithm. The offline training of MP-DNN is similar to R-DNN except that the $n$th training sample is $\left(\bar{\mathbf{r}}_{\mathcal{A},n},\frac{\mathbf{h}_{\mathcal{B},n}}{c}\right)$. For R-DNN and MP-DNN, the number of neurons of each layer is dependent on the ratio of high-resolution ADC antennas, $\eta$. The architectures of R-DNN and MP-DNN under some typical values of $\eta$ are presented in Table~\ref{table_2}.\footnote{We set different numbers of neurons in hidden layers for R-DNN and MP-DNN in different values of $\eta$ to match the corresponding input and output dimensions. This adaptive setting can balance the estimation accuracy and DNN complexity.}

\begin{table}[!t]
\centering
\caption{Architectures of the R-DNN and MP-DNN}
\label{table_2}
\begin{tabular}{p{1.5cm}<{\centering}|p{0.85cm}<{\centering}|p{0.30cm}<{\centering}|p{0.3cm}<{\centering}|p{0.85cm}<{\centering}|p{0.3cm}<{\centering}|p{0.3cm}<{\centering}|p{1.0cm}<{\centering}}
\hline
\multirow{2}{*}{Neural layer} & \multicolumn{3}{c|}{Size of R-DNN} & \multicolumn{3}{c|}{Size of MP-DNN} & \multirow{2}{*}{\makecell{Activation \\function}}\\
\cline{2-7}
 ~ & $\eta\!=\!0.2$ & $0.5$ & $0.8$ & $\eta\!=\!0.2$ & $0.5$ & $0.8$ & ~\\
\hline
 Input layer & 26 & 64 & 102 & 26 & 64 & 102 & - \\
\hline
 Dense \!layer \!1 & 50 & 120 & 200 & 50 & 120 & 200 & ReLU \\
\hline
 Dense \!layer \!2 & 100 & 200 & 400 & 100 & 200 & 100 & ReLU \\
\hline
 Dense \!layer \!3 & 50 & 120 & 200 & 140 & 120 & 50 & ReLU \\
\hline
 Output layer & 26 & 64 & 102 & 102 & 64 & 26 & tanh \\
\hline
\end{tabular}
\vspace{-0.3cm}
\end{table}

\vspace{-0.2cm}
\subsection{Summarize DI-DNN and SIP-DNN based Approaches}

For the channel estimation task in this article, the DI-DNN based approach can be regarded as the direct application of DNN. It simply incorporates the LS channel estimation of all antennas but neglects the adverse impact of the severely distorted signals quantized by the low-resolution ADCs on the estimation performance. In contrast, the SIP-DNN based approach is developed in a different philosophy to cover the shortage of the DI-DNN based approach by selectively utilizing the reliable observations corresponding to the high-resolution ADC antennas. The underlying idea is to establish a prediction mapping from the channels of high-resolution ADC antennas to those of low-resolution ADC antennas. We convert the original network to two parallel DNNs for ease of offline training. The SIP-DNN and DI-DNN based approaches have respective performance advantage and the combination of them makes the DNN based channel estimation framework quite sound for massive MIMO with mixed-ADC.

\vspace{-0.2cm}
\section{Numerical Results}

In this section, we evaluate the proposed DNN based channel estimation approaches by using numerical results. We set the number of antennas at the BS, $M=64$, the number of paths, $L=8$, the average power gain of each path $\sigma^2_{\alpha}=1$, the AoA, $\varphi_l$, is chosen randomly from the set of $\frac{2\pi}{20}\times\left[0,1,2,\ldots,19\right]$. The proposed DNNs are set as follows. The training set, validation set, and testing set contain $90,000$, $10,000$, and $10,000$ samples, respectively.\footnote{When generating the testing set, we use different AoAs from the training set. Therefore, the proposed DNNs are not the simple fitting on the specific training set but can learn the inherent channel structure and are suitable for different channel statistics.} The architecture of DI-DNN is detailed in Table~\ref{table_1} while SIP-DNN is set in Table~\ref{table_2} under some typical values of $\eta$. Adam is used as the optimizer. The number of epochs and learning rate are set as $100$ and $1\times10^{-3}$, respectively. The batch size is $128$. The scaling factor is set as $c=3$. The state of the art linear and nonlinear channel estimation methods are used for comparison: liner minimum mean-squared error (LMMSE) and expectation maximization Gaussian-mixture generalized approximate message passing (EM-GM-GAMP) in \cite{J. Mo_b}. According to \cite{P. Dong_a} and \cite{J. Mo_b}, the LMMSE estimator is given by $\hat{\mathbf{h}}_{\Delta,\textrm{LMMSE}}=\mathbf{C}_{\Delta}(\alpha\mathbf{C}_{\Delta}+\frac{\sigma_{0}^2}{P}\mathbf{I}_{|\Delta|}+(1-\alpha)\textrm{diag}( \mathbf{h}_{\Delta}\mathbf{h}_{\Delta}^H))^{-1}\bar{\mathbf{r}}_{\Delta}$, where $\Delta=\mathcal{A}$ or $\mathcal{B}$, $\mathbf{C}_{\Delta}=\mathbb{E}\{\mathbf{h}_{\Delta}\mathbf{h}_{\Delta}^H\}$ denotes the corresponding covariance matrix, the value of $\alpha$ is set as $0$ for $\Delta=\mathcal{A}$ and is taken from \cite[Table I]{P. Dong_a} according to the ADC resolution for $\Delta=\mathcal{B}$, respectively. To evaluate the channel estimation performance, we use the normalized MSE (NMSE), which is defined as $\textrm{NMSE}=\mathbb{E}\left\{\frac{\|\mathbf{h}-\hat{\mathbf{h}}\|^2}{\|\mathbf{h}\|^2}\right\}$ for the DI-DNN based approach and as $\textrm{NMSE}=\eta\mathbb{E}\left\{\frac{\|\mathbf{h}_{\mathcal{A}}-\hat{\mathbf{h}}_{\mathcal{A}}\|^2}{\|\mathbf{h}_{\mathcal{A}}\|^2}\right\} +(1-\eta)\mathbb{E}\left\{\frac{\|\mathbf{h}_{\mathcal{B}}-\hat{\mathbf{h}}_{\mathcal{B}}\|^2}{\|\mathbf{h}_{\mathcal{B}}\|^2}\right\}$ for LMMSE, EM-GM-GAMP, and the SIP-DNN based approach, respectively.

\begin{figure}[!t]
  \centering
  \subfigure[\scriptsize NMSE versus SNR with $\eta=0.5$]{\includegraphics[trim=10 0 0 0, width=2.65in]{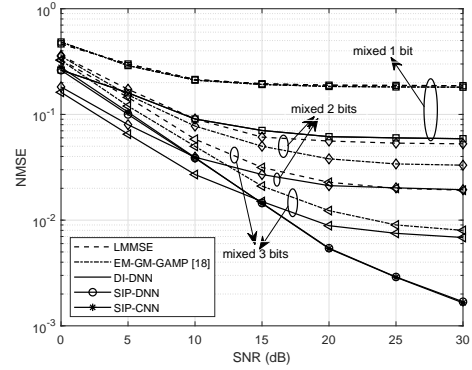}\label{nmse_block_snr}}
  \subfigure[\scriptsize NMSE versus $\eta$ with $\textrm{SNR}\!=\!20$ dB]{\includegraphics[trim=0 0 10 0, width=2.65in]{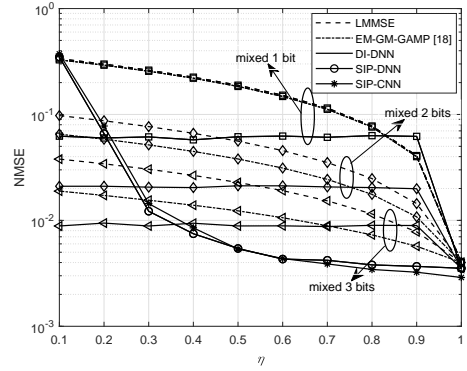}\label{nmse_block_eta}}
  \caption{NMSE for LMMSE, EM-GM-GAMP, and the proposed NN based approaches with block ADC resolution pattern.}
  \vspace{-0.4cm}
\end{figure}

Fig.~\ref{nmse_block_snr} shows the NMSE performance versus SNR for the LMMSE, EM-GM-GAMP, and proposed NN based approaches with $\eta=0.5$ and block ADC resolution pattern, where the high- and low-resolution ADC antennas are placed separately, i.e., $\mathcal{A}=\{1,2,\ldots,32\}$. From Fig.~\ref{nmse_block_snr}, the DI-DNN based approach always outperforms LMMSE and EM-GM-GAMP. However, their performance is limited by the error floor at the medium and high SNR regime due to the low-resolution ADCs in $\mathcal{B}$. In contrast, there is no significant error floor for the SIP-DNN based approach since it does not utilize those pilots quantized by low-resolution ADCs.\footnote{In the low SNR regime, the NMSE performance is limited by the SNR. When the SNR becomes high, the bottleneck turns to the number of training set data. So the slope of the SIP-DNN curve decreases as the SNR increases after 20 dB.} As a result, the SIP-DNN based approach outperforms LMMSE and EM-GM-GAMP at the whole SNR regime and the DI-DNN based approach at the medium and high SNR regime remarkably, especially with mixed one-bit ADCs.
The impact of $\eta$ on the NMSE performance is studied in Fig.~\ref{nmse_block_eta} with $\textrm{SNR}=20$ dB. The advantage of the DI-DNN based approach over LMMSE and EM-GM-GAMP is significant with a small $\eta$ and decreases along with the increase of $\eta$. The performance of the SIP-DNN based approach is poor when $\eta=0.1$ but is improved rapidly as $\eta$ increases. When $\eta\geq 0.4$, it achieves the best performance among all approaches. Interestingly, the points of intersection of the DI-DNN and SIP-DNN based approaches indicate that the system can select one of them to guarantee the best channel estimation performance according to the SNR or $\eta$. In addition, to evaluate the effect of different NN architectures on estimation accuracy, we use the CNN to replace the fully-connected DNN in SIP-DNN based approach, which is called a SIP-CNN in Fig.~\ref{nmse_block_snr} and Fig.~\ref{nmse_block_eta}. The CNN consists of the input layer, two convolutional layers, one flatten layer, one dense layer, and the output layer. It can be seen that the fully-connected and convolutional networks achieve very similar performance with different SNRs or $\eta$, which reveals that the basic fully-connected architecture is adequate for the channel prediction task.

\begin{figure}[!t]
  \centering
  \subfigure[\scriptsize NMSE versus SNR with $\eta=0.5$]{\includegraphics[trim=0 0 0 0, width=1.65in]{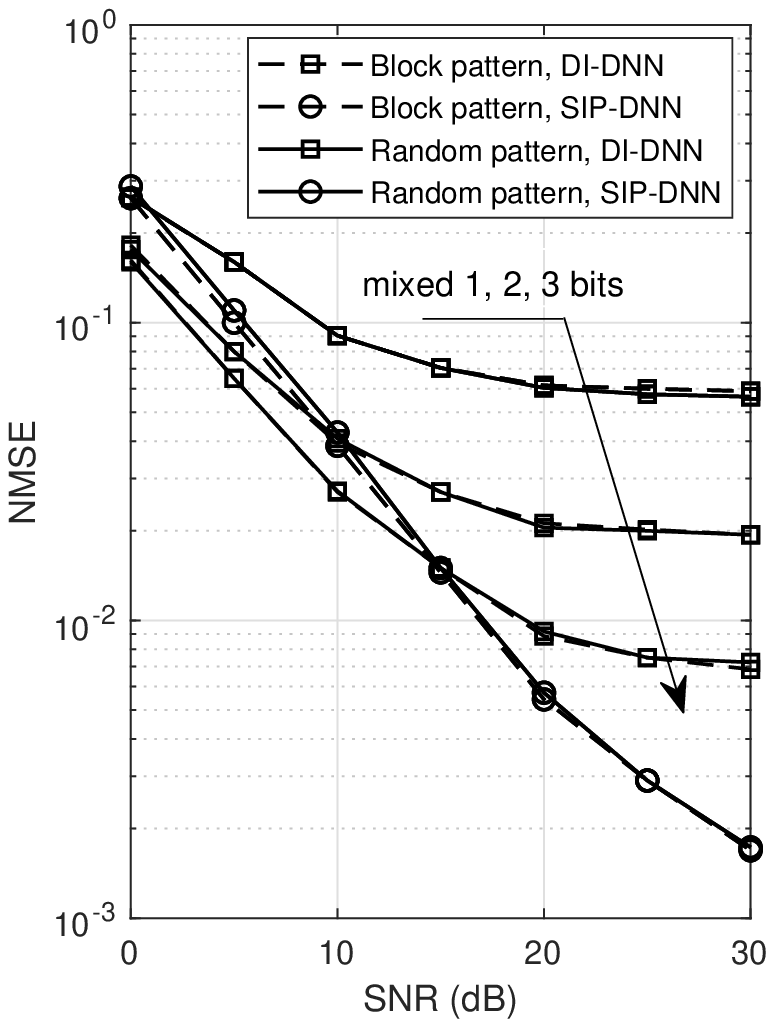}\label{nmse_rand_snr}}
  \subfigure[\scriptsize NMSE versus $\eta$ with $\textrm{SNR}\!=\!20$ dB]{\includegraphics[trim=0 0 0 0, width=1.65in]{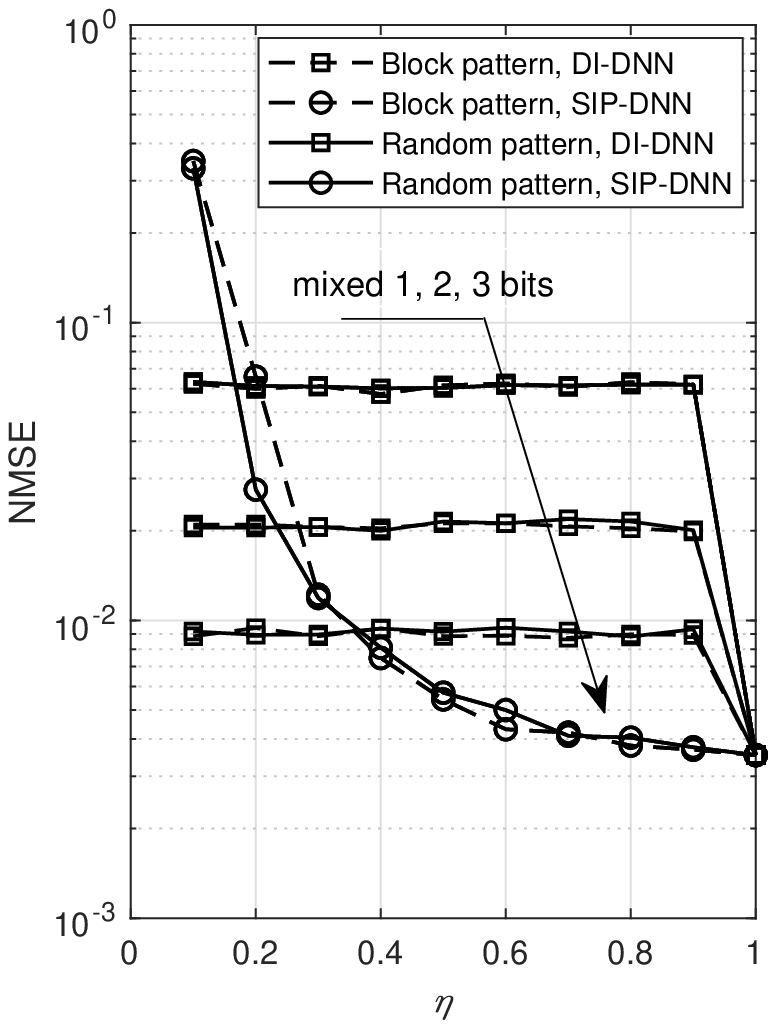}\label{nmse_rand_eta}}
  \caption{NMSE for the proposed DNN based approaches with block and random ADC resolution patterns.}\label{nmse_rand}
  \vspace{-0.4cm}
\end{figure}

Then we investigate a less structured random ADC resolution pattern where the antennas with high- and low-resolution ADCs are placed arbitrarily. Fig.~\ref{nmse_rand_snr} and Fig.~\ref{nmse_rand_eta} show the NMSE performance for the proposed DNN based approaches under block and random ADC resolution patterns versus SNR and $\eta$, respectively. From the figure, there is almost no performance loss even if the proposed approaches are applied to the unstructured ADC resolution pattern. Although the structured block ADC resolution pattern will be the mainstream configuration in future massive MIMO systems, Fig.~\ref{nmse_rand} is still important to demonstrate useful insights. It shows that the proposed DNN based approaches are robust to different ADC resolution patterns instead of dependent on the specific one. Furthermore, it reveals that the proposed approaches can learn the inherent spatial correlation in a quite comprehensive way and thus are promising to be widely applied to various types of ADC resolution patterns.

\section{Conclusion}

In this article, DL is adopted to address the challenging channel estimation problem in mixed-ADC massive MIMO systems. DI-DNN and SIP-DNN based approaches are developed to exploit the observations associated with all antennas and high-resolution ADC antennas, respectively, for channel estimation. Numerical results show that the proposed approaches are superior to the existing methods and are promising to be widely applied to practical systems with different ADC resolution patterns. The combination of the DI-DNN and SIP-DNN based approaches makes the DNN based channel estimation framework quite sound for massive MIMO with mixed-ADC.



\ifCLASSOPTIONcaptionsoff
  \newpage
\fi


\end{document}